\input epsf
\documentstyle{twocolprepr}
\twocolumn
\newcommand{\MFLG}{M.~F.~L.~Golterman}
\newcommand{\DNP}{D.~N.~Petcher}
\newcommand{\sect}[1]{{\bf \section{#1}}}
\newcommand{\be}{\begin{equation}}
\newcommand{\ee}{\end{equation}}
\newcommand{\bea}{\begin{eqnarray}}
\newcommand{\eea}{\end{eqnarray}}
\newcommand{\ea}{\end{array}}
\newcommand{\ba}{\begin{array}}
\newcommand{\nn}{\nonumber}
\newcommand{\eq}[1]{eq.~(\ref{eq:#1})}

\newcommand{\Sfermion}{{S_{\rm fermion}}}

\newcommand{\V}{V}
\newcommand{\Vdagger}{{\V^\dagger}}
\newcommand{\D}{{D}}
\newcommand{\Dtilde}{{\tilde{\D}}}
\newcommand{\Dmu}{{\D_\mu}}
\newcommand{\Dtildemu}{{\Dtilde_\mu}}
\newcommand{\DWmu}{{\D^{(W_\mu)}_\mu}}
\newcommand{\DWtildemu}{{\Dtilde^{(W_\mu)}_\mu}}
\newcommand{\latd}{{\partial}}
\newcommand{\latdtilde}{{\tilde \latd}}
\newcommand{\latdmu}{{\latd_\mu}}
\newcommand{\latdtildemu}{{\latdtilde_\mu}}
\newcommand{\PL}{{P_L}}
\newcommand{\PR}{{P_R}}
\newcommand{\tr}{{\rm tr}}
\newcommand{\hc}{\hbox{\rm h.c.}}
\newcommand{\gprime}{{g^\prime}}
\newcommand{\wprime}{w^\prime}
\newcommand{\muhat}{a_{\hat\mu}}
\newcommand{\nuhat}{a_{\hat\nu}}
\newcommand{\Umu}{{U_\mu}}
\newcommand{\Udaggermu}{{U^\dagger_\mu}}
\newcommand{\ULmu}{{U_{L\mu}}}
\newcommand{\UdaggerLmu}{{U^\dagger_{L\mu}}}
\newcommand{\URmu}{{U_{R\mu}}}
\newcommand{\UdaggerRmu}{{U^\dagger_{R\mu}}}
\newcommand{\ALmu}{{\vec A}_{L\mu}}
\newcommand{\Bmu}{B_\mu}
\newcommand{\Amu}{A_\mu}
\newcommand{\Wmu}{W_\mu}
\newcommand{\Wtildemu}{{\tilde{W}}_\mu}

\newcommand{\Wdaggermu}{W_\mu^\dagger}
\newcommand{\psibar}{{\overline{\psi}}}
\newcommand{\psiR}{\psi_R}
\newcommand{\psiL}{\psi_L}
\newcommand{\psibarL}{\psibar_L}

\newcommand{\psin}{\psi^{(n)}}
\newcommand{\psibarn}{\psibar^{(n)}}
\newcommand{\psiRn}{\psi_R^{(n)}}
\newcommand{\psiLn}{\psi_L^{(n)}}
\newcommand{\psibarLn}{\psibar_L^{(n)}}
\newcommand{\psibarRn}{\psibar_R^{(n)}}
\newcommand{\psic}{\psi^{(c)}}
\newcommand{\psiRc}{\psi_R^{(c)}}
\newcommand{\psiLc}{\psi_L^{(c)}}

\newcommand{\nuL}{{\nu_L}}
\newcommand{\nuR}{{\nu_R}}
\newcommand{\nubar}{{\overline\nu}}

\newcommand{\nubarR}{{\nubar_R}}
\newcommand{\epsilonR}{{\epsilon_R}}
\newcommand{\epsilonbar}{{\overline\epsilon}}
\newcommand{\epsilonbarR}{{\epsilonbar_R}}
\newcommand{\eL}{{e_L}}
\newcommand{\eR}{{e_R}}
\newcommand{\Phidagger}{{\Phi^\dagger}}
\newcommand{\Phitildemu}{{\tilde\Phi_\mu}}
\newcommand{\Fpropninv}{{S^{(n)-1}_F}}
\newcommand{\Fpropninvren}{S^{(n)-1}_{F\hbox{\rm ren}}}

\begin{document}
\title{{\bf Fermions in Models with Wilson--Yukawa Couplings}\thanks{Talk
presented by D.~Petcher at the Workshop on
Non-Perturbative Aspects of Chiral Gauge Theories, Rome, March, 1992}}

\preprint{Wash. U. HEP/92-81}
\date{June, 1992}
\author{Maarten F.L. Golterman and Donald N. Petcher}
\address{Department of Physics, Washington University, Saint Louis,
Missouri 63130, U.~S.~A.}

\begin{abstract}
Our work on models with Wilson--Yukawa couplings is reviewed. Conclusions
include the failure of such models to produce continuum chiral gauge
theories.
\end{abstract}

\maketitle

\sect{Introduction}

The need for a non-perturbative formulation of the standard model arises
primarily from two facts. First the upcoming generation of experiments
promises to press the limits where the perturbative
standard model will remain a good approximation to the actual situation.
But
more importantly, the triviality of the $\phi^4$ theory raises the question
what underlies the mechanism of symmetry breaking giving rise to masses
of the particles of the model. In view of this we are forced to think of the
standard model as an approximation to some yet unknown theory that
turns out to do a good job in fitting data in the relatively low
energy regime. The outstanding problem is to find a model (at first, any
model will do) that can reproduce the perturbative standard model and also
make some predictions concerning the non-perturbative situation. For lack of
other non-perturbative methods, many have turned to the lattice to search
for such a model, where the primary obstacle to the construction of chiral
fermions is the fermion doubling problem.

In this talk we review work in which we have been involved pertaining to the
model proposed by Smit and Swift \cite{SmitSwift} as a candidate for a
lattice standard model, and some variants thereof. This model seeks to
remove the fermion doublers by generating Wilson mass terms (or the effect
thereof) along with the normal mass terms. The terms added to the lagrangian
to accomplish this are generally called Wilson--Yukawa terms. As it turns
out, in order for the mechanism to work one must take the Wilson--Yukawa
coupling of the theory to be strong, which makes this inherently a
non-perturbative problem. We will discuss the phase diagram of such models,
and the mass spectrum for various regions of it. We will also discuss what
happens to the renormalized couplings so as to classify the interactions one
can obtain from such a model. Finally we will say a word about anomalies.
The conclusion of our work is sadly that these models do not appear to
contain a scaling region in which they agree with
the perturbative standard model.
\medskip
\sect{The Model}
The electron-neutrino sector of the
Smit--Swift proposal for a standard model on the lattice
(see e.g. ref. \cite{MGreview} and refs. therein)
is defined through the action
\bea
S &=&
\sum_{x,\mu}a^4\half\psibar(x)\gamma_\mu(\Dmu+\Dtildemu)\psi(x) \nn\\
&+&\sum_x a^4[\psibarL(x)\Phi(x)Y\psiR(x)+\hc]\label{eq:SSaction}\\
&-&\frac{w}{2}\sum_x a^4[\psibarL(x)\Phi(x)
a^2\Box\psiR(x)+\hc]\nn\\
&-&\frac{\kappa}{a^2}
\sum_{x,\mu}a^4\tr[\Phi(x)\ULmu(x)\Phi(x+\mu)\UdaggerRmu+\hc]\nn\\
&+&V(\tr\Phi\Phidagger)+\hbox{\rm pure gauge action}.\nn
\eea
This action produces the classical action for the
usual standard model as the lattice spacing $a$ is taken to
zero, with two exceptions:
fermion doublers are present for all the fermions, and a
right-handed neutrino is also built into the model. However,
the right-handed
neutrino decouples in the classical continuum limit.
With the addition of
this right-handed neutrino, the model can be written in a more symmetric
way than usual, with both the right-handed and the left-handed fermion
fields given in terms of the electron and neutrino fields via
\bea
\psiL = \left(\ba{c} \nuL \\ \eL\ea\right),\;
& \psiR = \left(\ba{c}\nuR\\\eR\ea\right).
\label{eq:fermionfields}
\eea
The lattice covariant derivatives on these fields appearing in the first
term of \eq{SSaction} is defined by
\bea
\Dmu\psi(x) &=&\frac{1}{a}(\Umu(x)\psi(x+\muhat)-\psi(x)),\\
\Dtildemu\psi(x) &=& \frac{1}{a}(\psi(x) - \Udaggermu(x-\muhat)),\nn
\eea
with the gauge fields given by link variables which are exponentials of
the $SU(2)\times U(1)$ Lie algebra elements $\ALmu$ and $\Bmu$:
\bea
\Umu &=& \ULmu \PL + \URmu \PR,\nn\\
\ULmu &=& e^{-iga\ALmu\cdot\frac{\vec\tau}{2}+i\half\gprime\Bmu},\\
\URmu &=& \left(\ba{cc}1&0\\0&e^{i\gprime\Bmu}\ea\right),\nn
\eea
indicating that the fermion fields transform as usual under the
$SU(2)\times U(1)$ chiral gauge symmetry, with the right-handed neutrino being
a singlet with respect to both groups.
$\Box$ is the covariant lattice laplacian:
\be
\Box = \sum_\mu\Dmu\Dtildemu.
\ee
In the second term, the Yukawa coupling is a matrix:
\bea
Y = \left(\ba{cc}0&0\\0&y_e\ea\right)
\eea
which couples electrons to the Higgs field $\Phi$ in the usual way;
this latter field can be parameterized by the product of an $SU(2)$ matrix
$\V$ and a positive function $\rho$:
\be
\Phi = \rho \V,\;\;\; \V\in SU(2).
\ee
The first two
terms in \eq{SSaction}
together form the fermion sector of the ``naive'' lattice standard
model, which in perturbation theory is exactly the
standard model except with doubled fermions (and the right-handed
neutrino).
The third term is added with the hope of curing the doubling disease. Note
that the right-handed neutrino couples to other fields only through this
term, which vanishes in the classical continuum limit.
The other fermionic sectors can be constructed similarly.

The fourth term is the kinetic term for the Higgs field which, along
with a single site term in the potential,
becomes the usual kinetic term proportional to $|\Dmu\Phi|^2$ in the
classical continuum limit.

Lastly we include the usual plaquette action for each of the gauge
fields, and a potential for the Higgs field. This potential term is not
necessary for our purposes and with the following justification
we will neglect it for the rest of the
paper.  Due to the triviality of $\phi^4$
theory, the pure scalar theory reproduces perturbative
$\phi^4$ theory in the scaling region for all values of the bare $\phi^4$
coupling, with the renormalized coupling vanishing in the strict continuum
limit. This being the case, we can equally well take the bare coupling to be
infinite, thereby freezing the Higgs field to a fixed radius. Thus we
suffer no loss in limiting ourselves
from now on to a theory with frozen Higgs field, for which $\rho$ is
replaced by a constant which we take to be one in lattice units:
\be
\Phi\stackrel{\lambda_{\phi^4}\to \infty}{\rightarrow}\frac{1}{a}\V,
\;\;\;\V\in SU(2).
\label{eq:freezePhi}
\ee
In this limit, the Higgs potential becomes an irrelevant constant that we
ignore. The lattice spacing appears in \eq{freezePhi} because $\Phi$
is a field of dimension $1$.

Aside from the $SU(2)\times U(1)$ chiral gauge symmetry, the
action in \eq{SSaction} is also invariant under a simple shift of the
right handed neutrino \cite{Usdecouple}:
\be
\nuR\rightarrow\nuR+\epsilonR,
\;\;\;\nubarR\rightarrow\nubarR+\epsilonbarR,
\label{eq:shiftsym}
\ee
a symmetry that will be useful in drawing some of our conclusions.

Finally, for pedagogical purposes,
we make one further simplification of the model.
Rather than discuss the full model as
described above, we limit ourselves to what could be called a
``generalized neutrino sector'' which contains only one left-handed
gauge symmetry,
that can be taken as either $SU(2)$ or $U(1)$. So the action we will
actually study is the following:
\bea
S &=&
\sum_{x,\mu}a^4\half\psibar(x)\gamma_\mu(\Dmu+\Dtildemu)\psi(x)
\label{eq:Simpleaction}\\
&+&\sum_x a^4[\psibarL(x)\Phi(x)(y-\frac{w}{2}a^2\Box)\psiR(x)+\hc]\nn\\
&-&\frac{\kappa}{a^2}
\sum_{x,\mu}a^4\tr[\Phidagger(x)\ULmu(x)\Phi(x+\mu)+\hc]\nn\\
&+&\hbox{\rm pure gauge action}\nn.
\eea
where now the laplacian
does not contain gauge fields because it
acts only on the gauge singlets $\Phidagger\psiL$ and $\psiR$. The model
defined through the action in \eq{Simpleaction} has
a local $SU_L(2)$ (or $U_L(1)$) gauge invariance and a rigid $SU_R(2)$ (or
$U_R(1)$) chiral invariance.  However, all results that we present here can
easily be generalized to the action given in \eq{SSaction}.

\sect{The Gauge--Higgs System}

Let us remind you first that for a continuum limit of a
lattice theory to be interesting, it
should be taken near a phase transition for which some correlation length
$\xi$ smoothly blows up in lattice units (second order or above).
Then consequently the excitations
in the lattice theory associated with the divergent
correlation length are the particles which survive, and their masses in
lattice units are vanishing:
\be
ma = \frac{1}{\xi} \rightarrow 0.
\ee
This is to say that just as in normal perturbation theory, we must tune the
couplings so as to insure
that the masses have no divergent part.
Our search is then
for some region in the phase diagram for which as the lattice
spacing is taken to zero, exactly the particles we would like to
see emerge as physical (i.e. with finite mass)
and all others' masses diverge (i.e. are of the order of the cutoff, $1/a$).
Let us first review what happens in the gauge--Higgs system.

In the pure (frozen) Higgs system, the large $\kappa$ region is a broken
phase for which the order parameter
\be
\langle \V(x)\rangle = av\ne 0,
\ee
so $v$ is a parameter that in general diverges with small lattice
spacing. For small $\kappa$ there is a symmetric phase in which
the order parameter $\langle \V\rangle$ vanishes identically.
There is a second order phase transition
in between for some critical value of $\kappa$ which we call $\kappa_c$. In
the `scaling region' near the phase transition where the renormalization
group can be applied to the long range physics, an effective perturbative
$\phi^4$ theory emerges \cite{MontvayJapan}.
To relate to the standard model, $v$ should be tuned to
the vacuum expectation
value of the Higgs field which is about $250$ GeV.

Now as the gauge fields are turned on, the critical point continues out into
the direction in the phase diagram of the gauge coupling \cite{FradkinShenker}.
Note that in unitary gauge
the Higgs kinetic term given in \eq{Simpleaction} becomes
\bea
\lefteqn{-\kappa\sum_{x,\mu}\tr[\ULmu(x)+\UdaggerLmu(x)] =}\\
&&\kappa g^2 \sum_{x,\mu}a^2\tr \ALmu^2(x) + O(a).\nn
\eea
This is therefore a mass term for the gauge field.
With the frozen Higgs field, one cannot do
renormalized perturbation theory in this gauge, but
if $\kappa$ is tuned to some $\kappa_c$ (which is non-zero due to quantum
fluctuations)
and renormalized perturbation theory should emerge, the tuning necessary for
this to occur is to insure that the mass of the gauge field does not
diverge.  As we are only after a theory which reproduces the standard model
perturbatively in the gauge couplings,
the perturbative mass of the gauge field (in lattice units) thus
becomes the effective order parameter. On the broken side of the phase
transition this is just the lattice realization of the perturbative
standard model.  On the symmetric side, one obtains a theory of $SU(2)$
scalar chromodynamics.

\sect{The Perturbative Solution with Fermions}

For the time being we turn off the gauge coupling completely to discuss the
fermion--Higgs sector of the model.
Now let us consider the case of small Yukawa couplings $y$ and $w$ in broken
phase of the Higgs theory.
{}From \eq{Simpleaction}, at tree level the Yukawa couplings
give rise to a momentum dependent mass term for the fermions
\be
M(p) = v\left(y-\frac{w}{2}\sum_\mu(1-\cos (ap_\mu))\right),
\ee
and as the lattice spacing goes to zero the poles in the propagator are
\be
m_{\rm fermion} = v(y+2nw), \;\;\; n = 0,\ldots,4.
\label{eq:weakcouplingmass}
\ee
Thus as usual in a theory of Wilson fermions,
there are $16$ fermions in the continuum spectrum.
\begin{figure}
\epsffile{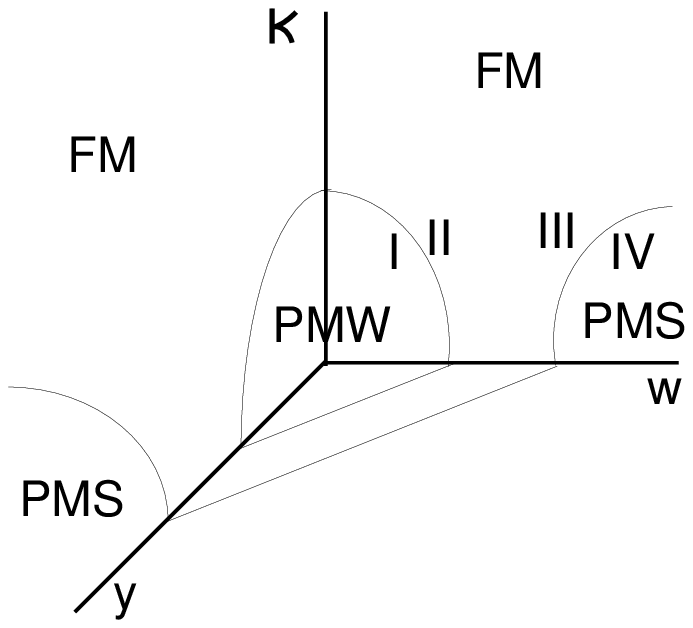}
\caption[phasediagram]{Phase diagram for the fermion--Higgs system.}
\end{figure}
Fig. 1 is a schematic representation of the phase diagram for the
Higgs--fermion model. When $y$ and $w$ are both small enough, $v$ vanishes,
and all fermions are degenerate and massless. This is the paramagnetic phase
in weak coupling (PMW phase) and is indicated by Roman numeral $I$ in the
phase diagram. As some combination of $y$, $w$ and $\kappa$ becomes large
enough, the theory crosses into a ferromagnetic (FM) phase, where the 16
fermions are non-degenerate according to \eq{weakcouplingmass}. This region
(for $y$ small) is indicated by Roman numeral $II$ in the phase diagram.
However, here as in the
perturbative standard model, all fermions have mass
proportional to $v$, including the doublers.
Hence in contrast to the case of QCD where the splitting between the
desired fermions and the doublers is of $O(\frac{1}{a})$,
here it is clearly not possible to remove the doublers from
the physical spectrum within the region of validity of perturbation theory
for the Yukawa couplings.
This means that if we hope to define a theory of
chiral fermions from the lattice, we are forced to consider a strong
coupling region, and the problem is inherently non-perturbative.

\sect{Beyond Perturbation Theory}

Before moving to strong coupling calculations, let us first introduce
fermion variables that are
singlets with respect to the chiral symmetry, viz.
\bea
\psiLn &=& a\Phidagger\psiL=\Vdagger\psiL,\\
\psiRn &=& \psiR.\nn
\eea
With this change of variables the fermion action becomes
\bea
\lefteqn{\Sfermion =}\label{eq:neutralaction}\nn\\
\lefteqn{\sum_{x,\mu}a^4\half\psibarLn(x)
\gamma_\mu(\DWmu+\DWtildemu)\psiLn(x)}\nn\\
\lefteqn{+\sum_{x,\mu}a^4\half\psibarRn(x)
\gamma_\mu(\latdmu+\latdtildemu)\psiRn(x)}\\
\lefteqn{+\sum_xa^4\psibarn(x)(y-\frac{w}{2}a^2\Box)\psin(x)}\nn\\
\lefteqn{-\frac{\kappa}{a^2}
\sum_{x,\mu}a^4\tr[\Phi(x)\ULmu(x)\Phi(x+\mu)+\hc]}\nn\\
\lefteqn{+\hbox{\rm pure gauge action,}\nn}
\eea
where
\bea
\DWmu\psiL(x) &=& \frac{1}{a}(\Wmu(x)\psiL(x+\muhat)-\psiL(x)),\\
\DWtildemu\psiL(x) &=&
\frac{1}{a}(\psiL(x)- \Wdaggermu(x-\muhat)\psiL(x-\muhat)),\nn
\eea
in which $\Wmu$ is the composite gauge field:
\be
\Wmu(x)=\Vdagger(x)\ULmu(x)\V(x+\muhat).
\ee
Thus all gauge couplings are now in the kinetic term of the left-handed
fermion.

Our first non-perturbative result is due to the shift symmetry in
\eq{shiftsym}. The Ward--Takahashi identities associated with this symmetry,
which coincide with the Schwinger--Dyson equations, gives the
following exact result for the form of the effective action,
\bea
\lefteqn{\Gamma(\psin,\psibarn,\Phi,\Phidagger,\ULmu)=}\\
&\sum_x a^4\{&\half\sum_\mu\psibarn\gamma_\mu(\latdmu+\latdtildemu)\psin\nn\\
&&+\psibarn(x)(y-\frac{w}{2}a^2\Box)\psin(x)\}\nn\\
&+\sum_{n>2}&\Gamma^{(n)}(\psiLn,\psibarLn,\Wmu),\nn
\label{eq:effectiveaction}
\eea
where $\Gamma^{(n)}$ is a one particle irreducible $n$-point function. Since
$\psiR$ appears only in $\Gamma^{(2)}$, all bare couplings of higher order
involving $\psiR$ vanish identically. It follows also that the exact inverse
propagator for the fermions, $\Gamma^{(2)}$, has the form
\bea
\lefteqn{\Fpropninv(p) }\nn\\
&=& \frac{i}{a}\sum_\mu\gamma_\mu(f_\mu(ap)\PL+\sin(ap_\mu)\PR)\\
&+&\frac{y}{a}+\frac{w}{a}\sum_\mu(1-\cos(ap_\mu)),\nn
\eea
where $f_\mu$ is determined by a dynamical calculation.  Actual dynamical
calculations indicate that $f_\mu$ is proportional to
$\sin(ap_\mu)$ \cite{JanFermi,Aachenfermion,Usoodpaper,Usoow,hoppingparam}.
Substituting this and renormalizing by multiplying
the bare field $\psin$ by $Z_2^{-\half}\PL+\PR$, where $Z_2$ is determined
from $f_\mu$, the renormalized propagator becomes
\bea
\Fpropninvren(p) &=& \frac{i}{a}\sum_\mu\gamma_\mu\sin(ap_\mu)\\
&+&\sqrt{Z_2}\left(\frac{y}{a}+\frac{w}{a}\sum_\mu(1-\cos(ap_\mu))\right)\nn.
\eea
Thus we see that mass renormalization is multiplicative, and
this propagator has the form of a free propagator for Wilson fermions.
As a consequence, if we set $y=0$, the right handed fermion completely
decouples as $a\to0$. It should also be emphasized that in general if we
had treated the full model represented by \eq{SSaction}, this would
guarantee the decoupling of the neutrino in that model, although the
right-handed electron would still couple as usual. These results clearly
indicate that $y$ must be tuned to obtain a non-zero physical mass for the
fermion.

Several strong coupling methods exist for calculating $Z_2$, including the
$1/d$ expansion \cite{Usoodpaper}, the $1/w$ expansion \cite{Usoow}, and the
hopping parameter expansion \cite{hoppingparam}, all of which are valid for
large values of $y+dw$ ($d$:spacetime dimension).
As we want to keep $y$ small so as to guarantee the
presence of a light fermion, we take $w$ large. Then we find
\bea
Z_2 &\sim & \frac{1}{z^2}\\
z^2&\equiv&\langle \Vdagger(x)\ULmu(x)\V(x+\muhat)\rangle,\nn
\eea
where $z^2$ is simply a number which is smooth and non-vanishing
across the phase transition
boundary. In the $1/d$ expansion this is seen explicitly:
\be
Z_2 \sim 2d.
\ee
Thus the mass spectrum for the neutral fermion
in the strong coupling region is given by
\be
am^{(n)}_{\rm fermion} = \sqrt{Z_2}(y+2nw),\;\;\;n=0,\ldots,4,
\ee
and if $y$ is tuned to zero as
\be
y\rightarrow a{\overline m} + O(a^2),
\ee
while $w$ is held finite, one light fermion remains while all doubler masses
are of the order of the cutoff. This occurs in regions $III$ and $IV$ of the
phase diagram in fig. 1.
This should be compared to \eq{weakcouplingmass} where
all masses are proportional to the vacuum expectation value and the doublers
do not decouple. Note that in contrast to the PMW phase, the fermions are
typically massive in the PMS phase, and the mechanism by which they get
their mass is distinctly different from the Higgs mechanism.

We have thus shown that the neutral fermion appears in the spectrum in the
strong coupling region, without doublers. We can also define a charged
fermion
\bea
\psiLc &=& \psiL,\\
\psiRc &=& a\Phi\psiR = \V\psiR.\nn
\eea
and ask whether this particle appears in the spectrum.
In the broken phase, the charged
fermion is simply related to the neutral fermion
\be
\psic \sim av\psin,
\label{eq:psicvanishing}
\ee
so that these two operators produce the same states in the Hilbert space.
In
the symmetric phase, we have examined the propagator of the charged field in
the $1/w$ expansion \cite{Usoow}, with the conclusion that it does not
contain a pole in momentum space, and thus the charged fermion does not have
a particle interpretation in the strong coupling symmetric (PMS) phase. This is
corroborated by Monte Carlo data \cite{Aachenoow,Dethisconf}.

\sect{Fermion Interactions}

Now that we have established the mass spectrum of the Smit--Swift model in
the strong coupling region and it has passed the two tests that the
right-handed neutrino decouples and the doublers also decouple, we have to
go on to examine the couplings of these particles with other fields. First
we examine the Yukawa couplings. From \eq{effectiveaction} we have already
seen that all bare couplings to the right-handed (neutral) fermion vanish.
In particular, this means that the Yukawa coupling vanishes. Let us now
examine the left-handed fermion. With the gauge
fields turned off in the $U(1)$ theory,
on general symmetry principles one can establish that the
scalar coupling to the left-handed fermion has the following form:
\bea
\lefteqn{\Gamma^{(3)}(\psiLn(p),\psibarLn(q),\phi_i(k))=}\\
&&(2\pi)^4\delta(p-q+k)\gamma_\mu\PL V^i_\mu(p,q),\;\;i=1,2,\nn
\eea
where $\Phi = \phi_1+i\phi_2$ is the renormalized boson field,
and $V^i_\mu$ is a dimensionless function. In
the symmetric phase, $V^i_\mu$ vanishes because this three point function is
not chirally symmetric. Therefore in the broken phase near the phase
transition it must be
proportional to the order parameter $\langle\V\rangle = av$ so as to preserve
continuity across the phase transition. Since the renormalization
coefficients are finite numbers, this Yukawa coupling vanishes with the
lattice spacing in the scaling region.  We have explicitly checked this in
the $1/d$ expansion \cite{Uscouplings}.

A similar conclusion can be
reached for higher order couplings to the neutral fermion \cite{Uscouplings}.
Thus we are left with a neutral fermion that also decouples entirely from the
Higgs field, and consequently there is no direct interaction between the
Higgs field and the fermions at all.

What of the gauge interaction? From \eq{neutralaction}, we might think that
the left-handed neutral fermion could couple to the composite gauge field
$\Wmu$, with no coupling to the right-handed fermion.
However a dimensional argument indicates that this coupling also
vanishes: from \eq{neutralaction} the gauge coupling term in the action is
\bea
\lefteqn{S_{\rm int}
\propto\sum_x a^4\psibarLn\gamma_\mu(x)
(\Wmu(x)\psiLn(x+\muhat)}\nn \\
&&\;\;\;\;\;\;\;- \Wdaggermu(x-\muhat)\psiLn(x-\muhat))\\
\lefteqn{=\sum_x a^4\psibarLn(x)\gamma_\mu
a^2(\Phidagger(x)\Dmu\Phi(x)-{\rm \hc})\psiLn(x),}\nn
\eea
so that the coupling takes place through the operator
$\Wtildemu=\Phidagger(x)\Dmu\Phi(x)-{\rm \hc}$,
which obviously has dimension three.
To relate it to a continuum gauge
field of dimension one, we need the square of
a dimensional parameter that comes out of the theory. In the
broken phase this parameter is $v^2$ whereas in the symmetric phase it
is the $\Lambda$ parameter of the gauge theory, a la $\Lambda_{QCD}$:
\be
\ba{ccll}
\Wtildemu&\propto& v^2 A^{(cont)}_\mu,&\hbox{\rm broken phase},\\
\Wtildemu&\propto& \Lambda^2 A^{(cont)}_\mu,&\hbox{\rm symmetric phase}.
\ea
\ee
This leads to a continuum form of the interaction term (in the broken phase)
\be
S_{\rm int}\stackrel{a\to0}{\rightarrow}
\propto a^2 v^2\int d^4x\;\psibarLn\gamma_\mu
A^{(cont)}_\mu\psiLn,
\ee
which vanishes in the limit. Thus the left-handed fermion not only decouples
from the Higgs field, but it also decouples from the gauge field as well,
and we are left with only a gauge--Higgs system and free fermions.

This argument can be substantiated through an explicit calculation in strong
Wilson--Yukawa coupling.
Since $\Wtildemu$ and $\Amu$ have the same quantum numbers on
the broken side of the phase transition, we can probe the fermion
gauge coupling by using the gauge field $\Amu$.
Treating this gauge field as a background field (i.e.
expanding in small gauge coupling), in a hopping parameter expansion for the
fermions but in which the Higgs field is treated exactly, one can show that
the one particle irreducible gauge coupling to the fermions is
proportional to $a^2$ times
the vacuum polarization of the gauge field for the pure
gauge--Higgs system up to lattice corrections that vanish in the continuum
limit. Since $a^2$ times
the mass of the vector boson is the order parameter in this
system, the gauge coupling also must vanish accordingly.
Here we report the results of this calculation.
To first order in the gauge coupling the one particle irreducible two point
function for the gauge field in a gauge--Higgs system is given by
\bea
\Gamma^{(2)}(\Amu(p)A_\nu(q))
=-g^2\kappa\delta(p+q){\cal P}_{\mu\nu}(p),
\eea
where
\bea
\lefteqn{{\cal P}_{\mu\nu}(p)
=\frac{1}{a^2}\delta_{\mu\nu}\langle\Vdagger(x)\V(x+\muhat)+\hc\rangle}
\label{eq:vacpol}\\
&+&\frac{\kappa}{a^2}\sum_z e^{-ipz}\frac{1}{a^2}\nn\\
\lefteqn{\times\langle(\Vdagger(0)\V(\muhat)-\hc)
(\Vdagger(z)\V(z+\nuhat)-\hc)\rangle.}\nn
\eea
Calculating the one particle irreducible three point function for the gauge
interaction to first order in the gauge coupling and hopping parameter
expansion ($\alpha = 1/(4w+y)$) in the theory with fermions we find
\bea
\lefteqn{\Gamma^{(3)}(\psin(p),\psibarn(q),A_\nu(k))}\\
&&= \half\alpha^2 g\delta(-p+q-k)e^{i\half a k_\nu}\gamma_\mu\PL
{\cal G}_{\mu\nu}(p,k)\nn\\
&&+ O(\alpha^4),\nn
\eea
where
\bea
\lefteqn{{\cal G}_{\mu\nu}(p,k)}\nn\\
&=& e^{ia(p_\mu+\half k_\mu)}\delta_{\mu\nu}\langle
\Vdagger(0)\V(\muhat)\rangle+\hc\\
&+&\sum_z e^{-ikz}\lbrack e^{ia(p_\mu+\half k_\mu)}\nn\\
&&\times\langle
\Vdagger(0)\V(\muhat)\left(\Vdagger(z)\V(z+\nuhat)-\hc\right)\rangle\nn\\
&&\;\;\;\;\;\;\;\;+\hc\rbrack.\nn
\eea
{}From these results one easily finds
\bea
\lefteqn{{\cal G}_{\mu\nu}(p,k) = a^{2}
\cos(a p_\mu+\half a k_\mu){\cal P}_{\mu\nu}(p)}\label{eq:GtoPrel}\\
&+&i\sin(a p_\mu+\half a k_\mu)\kappa{\cal Q}_{\mu\nu}(ak)\nn
\eea
where
\bea
\lefteqn{{\cal
Q}_{\mu\nu}(ak)=\sum_z e^{-ikz}\langle(\Vdagger(\muhat)\V(0)
+\hc)}\nn\\
&&\times(\Vdagger(z)\V(z+\nuhat)-\hc)\rangle.
\eea
In general the vacuum polarization \eq{vacpol} diverges quadratically.
Only at the critical value of $\kappa$, which by definition is the point
where the quadratic divergence vanishes, does the gauge boson appear in the
physical spectrum. Thus the first term in \eq{GtoPrel} vanishes as $a^2$.
The second
term can be easily shown to vanish as $a^2$ also, resulting in a
vanishing gauge coupling. Thus our previous dimensional argument for the
vanishing of the gauge coupling is demonstrated explicitly to first order in
the $1/w$ expansion, in which the scalar sector has been treated
exactly.

Note that mean field theory for the scalar field would
have given us the same result, but the  importance of the above calculation
lies in the fact that mean field theory can be misleading. Take for example the
case of the
fermion mass.  In that case if one only uses mean field theory to determine the
effects of the scalars, the resulting fermion propagator is sick. However,
this is repaired even if only the first order term in the $1/d$ expansion is
included \cite{Usoodpaper}.
In the present case however, mean field theory turns out to
be qualitatively correct as demonstrated in the above computation.

\sect{A Modified Smit--Swift Model}

As we have seen, not only do the doublers and the right-handed neutrino
decouple in the scaling region for strong $w$, {\it all fermions decouple}.
One may wonder if this is a necessary consequence of the model, or if it can
be formulated differently so as to restore the interactions. To address this
question we have studied the same model with the following slight
alteration. In place of the Yukawa terms given in the lagrangian in the
second line of \eq{Simpleaction}, we substitute
\be
\sum_x a^4\left[\psibarL(x)(y-\frac{\wprime}{2}
a^2\Box)\Phi(x)\psiR(x)+\hc\right].
\label{eq:wprimeterm}
\ee
The only difference between this term and the previous one is that $\Phi$
has now been moved from the site of $\psiL$ to that of $\psiR$. This has the
effect of restoring the necessity to include gauge fields in the laplacian
because
neither $\psi_L$ nor the combination $a\Phi\psiR$ is a singlet under gauge
transformations. In the strong coupling calculation, now the field in which
we must expand is a charged field so the role of the charged fermion and the
neutral fermion are interchanged in that region. Thus the
physical fermion is now charged and the neutral fermion does not appear in
the spectrum. Since a charged fermion is now physical, the remaining
question is whether this fermion will couple chirally to the other fields.
The answer is alas no. In this case the neutral right-handed fermion forms a
bound state with the Higgs field to provide a partner to the left-handed
fermion, forming a Dirac field. This subsequently couples in a vector-like
fashion to the gauge fields \cite{Uscouplings}.

If terms of both the type appearing in the original lagrangian and the
type in \eq{wprimeterm} are included, then which field
appears in the physical spectrum is determined by the relative strength of
the couplings $w$ and $\wprime$. If $w>\wprime$, the neutral particle is
physical and if $w<\wprime$ the charged particle is physical. For the two
couplings equal, the two particles both appear in the physical spectrum and
are degenerate. Fermion-gauge couplings are always vector-like.
For further discussion on this point see ref.
\cite{Usoowtsukuba}.

\sect{Anomalies}

One might be surprised to see that gauge
anomalies seem to play no role whatsoever
in the failure of Smit--Swift-like models to produce a continuum theory of
chiral fermions. Yet this appears to be correct. All our arguments deal
directly with propagators and couplings, and no triangle
diagrams play any role in the discussion. Note in this context that the
Eichten--Preskill model \cite{EP},
which pays careful attention to anomalies, seems to
have the same problem as the models considered here \cite{UsEP}.


There is also another potential problem related to the model as raised
recently by Banks \cite{Banks,Banks2},
stemming from the fact that the model has
an explicit global $U(1)$ symmetry which on the lattice is
exactly conserved, yet in the actual standard model this symmetry is
anomalous. Although one may attempt to get around this problem by adding
terms that explicitly break all symmetries that are not preserved by the
target theory as in the philosophy of Eichten and
Preskill \cite{EP,UsEP}, we feel that another scenario is possible
which is not covered by the illustrative models offered by Banks and
Dabholkar \cite{Banks2},
but rather coincides with the scenario offered by Dugan and
Manohar \cite{DuMan}.  Although because of the failure of the Smit--Swift
model to produce chiral fermions this scenario is not realized, it is
nevertheless instructive, so we would like to explain it here.

Consider the Noether current for the global $U(1)$ symmetry in the model of
\eq{Simpleaction}:
\bea
\lefteqn{{\cal J}_\mu^{\rm Noether} = }\\
\half\psibar(x)\gamma_\mu(\ULmu(x)\PL+\PR)\psi(x+\muhat) + \hc\nn\\
\lefteqn{-\half
w\lbrack\psibar(x)\Phitildemu(x)\psi(x+\muhat)
-\hc\rbrack,}\nn
\eea
with
\be
\Phitildemu(x) = (\Phidagger(x+\muhat)\PL+\Phi(x)\PR).
\ee
This Noether current is also explicitly gauge invariant because the gauge
symmetry commutes with the $U(1)$ symmetry. However, as we have shown, the
physical states in the strong coupling region turn out to be the neutral
fermions $\psin$. Thus if we transform to these variables, the Noether
current becomes
\bea
{\cal J}_\mu^{\rm Noether} &=& \half\lbrack\psibarLn(x)
\Wmu(x)\gamma_\mu\psiLn(x+\muhat)\nn\\
&+& \psibarRn(x+\muhat)\gamma_\mu\psiRn(x)\rbrack+ \hc
\label{eq:currentneutral} \\
&+&\frac{w}{a}\lbrack\psibar(x+\muhat)\psin(x)-\hc\rbrack.\nn
\eea
However, in terms of these variables, the kinetic term of the
action in \eq{neutralaction} clearly has the form of a term with a chiral
gauge invariance in terms of the composite field $\Wmu$
under which only the left-handed field transforms.
However this invariance is broken by the Wilson term in that action that
comes after the change of variables
from the original Wilson--Yukawa term. Although this `residual' gauge
invariance is not the original gauge invariance of the theory, a case can be
made as to why it is the appropriate gauge invariance to identify with that
of a continuum theory.  The physical states of a system represented by
gauge invariant states on the lattice, can be identified with corresponding
states in the continuum as written in unitary gauge. Using this rule, since
the physical fermion state emerging from the lattice theory is $\psin$, this
should be identified with the fermion of the continuum theory, and likewise
its associated gauge field $\Wmu$ can be identified with the continuum gauge
field (in unitary gauge they are the same).  Thus returning to the Noether
current in \eq{currentneutral} in terms of the neutral fermions, this
current is not invariant under the `residual' gauge invariance associated
with the composite field $\Wmu$. A gauge invariant current can be formed by
removing the term proportional to $w$, but this is no longer the $U(1)$
Noether current. Thus the term proportional to $w$ builds up the anomaly
associated with the `residual' gauge invariance in just the same way as it
does in QCD, as demonstrated by Karsten and Smit \cite{KarSmit}. This
scenario coincides exactly with the one outlined by Dugan and Manohar.
However, we should emphasize that as far as we know, the scenario is not
realized in any model, although it remains a distinct possibility.
In all models similar to those studied
here, the fermions form vector representations, thus avoiding the problem.
\bigskip
\acknowledge
We would like to thank the participants of this meeting, in particular
L.~Randall, for discussions. We also would like acknowledge J.~Smit and
E.~Rivas with whom a part of the work reported here was done. Part of this
work was supported by the D.~O.~E. We would also like to thank the
organizers and participants of this workshop for a very lively time of
discussion and a most enjoyable stay in Rome.

\newcommand{\NPB}[1]{Nucl. Phys. {\bf B#1}}
\newcommand{\NPBP}[1]{Nucl. Phys. {\bf B} (Proc. Suppl.) {\bf #1}}
\newcommand{\PRD}[1]{Phys. Rev. {\bf D#1}}
\newcommand{\PLB}[1]{Phys. Lett. {\bf #1B}}

\end{document}